\documentclass[conference]{IEEEtran}
\IEEEoverridecommandlockouts
\usepackage{amsmath,amssymb,mathtools}
\usepackage{cite}
\usepackage{graphicx,booktabs,array,xcolor}
\usepackage{microtype}
\usepackage[colorlinks=true,linkcolor=blue,citecolor=blue,urlcolor=blue]{hyperref}
\newcommand{\Ca}{C_\alpha}
\newcommand{\dd}{\,\mathrm{d}}
\newcommand{\SID}{\operatorname{SID}}
\newcommand{\erfc}{\operatorname{erfc}}
\newcommand{\ind}{\boldsymbol{1}}
\newtheorem{proposition}{Proposition}
\newcommand{\um}{\mu\mathrm{m}}
\hypersetup{pdftitle={Exact Cap Response and Aperture Selection for Molecular Communication under Aligned Drift},pdfauthor={Yen-Chi Lee, Chun-An Yang, Nan Yang}}

\title{Exact Cap Response and Aperture Selection for Molecular Communication under Aligned Drift%
\thanks{This work was supported by the National Science and Technology Council of Taiwan (NSTC 113-2115-M-008-013-MY3).}}
\author{\IEEEauthorblockN{Yen-Chi Lee\IEEEauthorrefmark{1}, Chun-An Yang\IEEEauthorrefmark{2}, and Nan Yang\IEEEauthorrefmark{3}}
\IEEEauthorblockA{\IEEEauthorrefmark{1}Department of Mathematics, National Central University, Taoyuan, Taiwan}
\IEEEauthorblockA{\IEEEauthorrefmark{2}Department of Computer Science and Information Engineering, National Taiwan Normal University, Taipei, Taiwan}
\IEEEauthorblockA{\IEEEauthorrefmark{3}School of Engineering, The Australian National University, Canberra, Australia}
\IEEEauthorblockA{Emails: yclee@math.ncu.edu.tw; cayang@ntnu.edu.tw; nan.yang@anu.edu.au}}

\begin{document}
\maketitle
\begin{abstract}
Spatial reception under directed molecule transport requires balancing useful molecule capture against intersymbol interference (ISI). To characterize this balance, we study a fully absorbing spherical receiver that counts absorption events only within a prescribed cap. Using the known joint first-hitting law under aligned drift, we derive exact cap impulse, cumulative, and eventual responses. We perform stable cumulative evaluation. We then assess the derived responses through particle simulations and time-step comparisons, and demonstrate that our conditional-timing analysis explains when opposite drifts of equal magnitude can favor the same wider aperture. We also show that for the studied geometry and a symbol duration of $0.15$ s, the aperture that maximizes the signal-to-interference difference (SID) increases from $29.13^\circ$ at zero drift to $61.86^\circ$ at a drift magnitude of $40\,\mu\mathrm{m}$/s. These results (i) connect joint arrival-time and location statistics to spatial reception design, (ii) characterize the tradeoff between useful molecule capture and ISI, (iii) support SID-based aperture selection for a given drift and symbol duration, and (iv) quantify the SID loss incurred by retaining the zero-drift optimal aperture for the same symbol duration.
\end{abstract}


\section{Introduction}
\label{sec:introduction}
\begingroup
\color{black}
\ifdefined\hypersetup
  \hypersetup{citecolor=black,linkcolor=black,urlcolor=black}
\fi

Diffusion-based molecular communication conveys information through molecular propagation at microscales \cite{akyildiz2008,farsad2016,jamali2019,yeh2012new}. Late-arriving molecules create channel memory, requiring a balance between timely molecule capture and intersymbol interference (ISI). Directed transport further complicates this balance: drift can accelerate the delivery of useful molecules while carrying residual molecules away from the receiver, affecting both the desired signal and interference \cite{noel2014flows}. Moreover, drift redistributes both useful and late-arriving molecules over the receiver surface. Compared with full-surface counting, a suitably selected cap at the receiver improves the signal-interference balance by excluding regions contributing more late-arriving molecules than useful ones. Thus, counting region design must jointly account for arrival time and spatial distribution of absorbed molecules.

A fully absorbing spherical receiver that counts molecules only within a transmitter-facing cap was investigated in \cite{akdeniz2018}, where the signal-to-interference difference (SID) and bit-error-rate (BER) analysis employed an approximate zero-drift response. The associated compensation factors were derived using zero-drift time/angle marginals and thus did not directly cover nonzero drift. This raises two questions: how should the counting cap be selected under signed aligned drift, and what is the cost to retain the zero-drift aperture for the same symbol duration? Existing full-surface responses under zero drift \cite{yilmaz2014} and uniform drift \cite{lee2026} characterized total absorption but did not address spatial selection. To bridge this gap, we derive exact cap-restricted responses from the joint arrival-time and absorption-location law under signed aligned drift, retaining the temporal-spatial dependence needed for aperture selection.

Our cap-restricted response analysis builds on established joint first-hitting laws. Specifically, \cite{lee2026} employed the joint density and a drift tilt to obtain a full-surface channel impulse response (CIR). \cite{yin2009} derived the joint arrival-time and absorption-location densities for both drift-free and drifted diffusion outside a sphere. Despite such advances, we note that uniform aligned drift is a controlled model of directional transport, and more complex settings, such as receiver-disturbed flow or spatially varying velocities, require separate treatment \cite{jamali2019}.


Our main contributions are: i) we derive the cap-restricted CIR and cumulative distribution function (CDF), with the cumulative response evaluated by subtracting damped tails from infinite-time values to avoid an undamped spectral integral, ii) we validate the derived results through particle simulations using the same model parameters, including time-step comparisons based on shared Brownian paths, and iii) we analyze the conditional SID/aperture and conduct a three-duration study to quantify the cost of retaining the zero-drift design for each symbol duration. Crucially, conditional timing and absolute capture are distinct: under the stated condition, opposite drifts of equal magnitude can favor the same aperture, but yield different adaptation payoffs.



We clarify that the cap at the receiver determines which absorption events are counted, while its entire spherical surface remains absorbing. Thus, molecules hitting the surface outside the cap are permanently removed. Receivers with heterogeneous receptor coverage \cite{huang2023} and molecule-harvesting transmitter surfaces \cite{huang2025} involve different reaction-boundary dynamics. Our model therefore provides a tractable baseline for studying broader receiver reception and surface-reaction, rather than heterogeneous reaction kinetics.


\par
\endgroup

\section{System Model}
\label{sec:model}
\begin{table}[t]
\caption{Principal Notations}
\label{tab:notation}
\centering\footnotesize
\setlength{\tabcolsep}{3pt}
\renewcommand{\arraystretch}{1.04}
\begin{tabular}{@{}>{\raggedright\arraybackslash}p{.22\columnwidth}>{\raggedright\arraybackslash}p{.74\columnwidth}@{}}
\toprule
Symbol & Meaning \\
\midrule
$D,\ v$ & Diffusion coefficient; signed drift velocity ($v<0$ toward, $v>0$ away from receiver). \\
$\tau,\ \Theta$ & Time and polar angle of the first hit. \\
$\rho,\ \beta,\ b$ & $r_0/r$, $vr/(2D)$, and $|\beta|$ (dimensionless). \\
$\kappa$ & $v^2/(4D)$; the rate in $e^{-\kappa t}$. \\
$q_v,\ g_v$ & Joint hitting densities with respect to time--surface area and time--polar angle, respectively. \\
$h_v(t;\alpha)$ & Cap CIR (inverse time). \\
$H_v(t;\alpha)$ & Cap absorption probability by time $t$; unconditioned on absorption. \\
$H_v(\infty;\alpha)$ & Eventual cap absorption probability; unconditioned on absorption. \\
$T_s$ & Symbol duration. \\
$S_v,\ I_v$ & Counted probabilities of molecules arriving within the symbol interval and at all later times, respectively. \\
$\SID_v(\alpha)$ & $S_v(\alpha)-I_v(\alpha)$; Signal-to-interference difference. \\
$\alpha_v^{\ast},\ \alpha_0^{\ast}$ & SID-maximizing apertures at drift $v$ and zero drift for the same $T_s$. \\
$L_{\mathrm{fixed}}(v)$ & SID loss from retaining $\alpha_0^{\ast}$ under drift $v$. \\
\bottomrule
\end{tabular}
\end{table}

We consider a molecular communication system where the fully absorbing receiver is a sphere of radius $r$ centered at the origin, and all key notations used in system modelling are summarized in Table \ref{tab:notation}. A point transmitter is at $\boldsymbol{x}_0=r_0\boldsymbol{k}$, where $r_0>r$ and $\boldsymbol{k}$ is a fixed unit vector. In an unbounded three-dimensional medium, the position of each molecule at time $t$ is represented by 
$\boldsymbol{X}_t\in\mathbb{R}^3$. It evolves as $\dd\boldsymbol{X}_t=v\boldsymbol{k}\dd t+\sqrt{2D}\dd\boldsymbol{B}_t$ with $\boldsymbol{X}_0=r_0\boldsymbol{k}$, where $D>0$ is the diffusion coefficient and $\boldsymbol{B}_t$ is the standard three-dimensional Brownian motion. A negative signed velocity $v$ points toward the receiver; a positive $v$ points away from it.

\begin{figure}[t]
\centering
\includegraphics[width=.64\columnwidth]{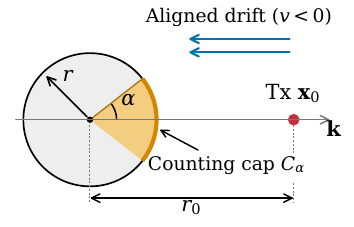}
\caption{Cross-sectional schematic of a fully absorbing sphere with a counted cap facing the transmitter. The illustrated drift points toward the receiver ($v<0$), although either sign of $v$ is allowed.}
\label{fig:system}
\end{figure}

Let $\tau=\inf\{t>0:\|\boldsymbol{X}_t\|=r\}$ denote the global first-hitting time, where $\|\cdot\|$ is the Euclidean norm on $\mathbb{R}^3$, and let $\boldsymbol{Y}=\boldsymbol{X}_\tau$ on $\{\tau<\infty\}$. The polar angle $\Theta=\angle(\boldsymbol{k},\boldsymbol{Y})$ determines whether the hit is counted. The counting cap is $\Ca=\{\boldsymbol{y}:\|\boldsymbol{y}\|=r,\ \angle(\boldsymbol{k},\boldsymbol{y})\leq\alpha\}$, with $0\leq\alpha\leq\pi$. Fig. \ref{fig:system} illustrates this geometry and the transmitter-facing counting cap. Every surface hit terminates a trajectory, including hits outside $\Ca$. Thus, $\alpha$ selects recorded events without changing the absorbing boundary condition.

We denote $q_v(t,\boldsymbol{y})$ as the joint density with respect to time and physical surface area $\dd S(\boldsymbol{y})$. For any measurable region $A$ on the receiver surface and any finite $t\geq0$, we obtain
\begin{equation}\label{Eq:Prob_1}
\mathbb{P}(\tau\leq t,\boldsymbol{Y}\in A)
=\int_0^t\int_A q_v(s,\boldsymbol{y})\,\dd S(\boldsymbol{y})\,\dd s.
\end{equation}
This density is not conditioned on eventual absorption; its integral over all times and the entire receiver surface equals the eventual hitting probability, which may be less than one.

With azimuthal symmetry, the joint time-polar-angle density is $g_v(t,\theta)=2\pi r^2\sin\theta\,q_v(t,\boldsymbol{y})$,
where $\boldsymbol{y}$ is any point on the receiver surface with polar angle $\angle(\boldsymbol{k},\boldsymbol{y})=\theta$.
The cap CIR and cumulative response are $h_v(t;\alpha)=\int_0^\alpha g_v(t,\theta)\dd\theta$ and $H_v(t;\alpha)=\mathbb{P}(\tau\leq t,\Theta\leq\alpha)=\int_0^t h_v(s;\alpha)\dd s$, respectively. The density $h_v$ has units of inverse time, while $H_v$ is a dimensionless counted fraction. We clarify that $H_v(t;\alpha)$ is not conditioned on eventual absorption, i.e., nonhitting molecules contribute zero. We denote its infinite-time limit by $H_v(\infty;\alpha)$.

\section{Exact Cap Response under Aligned Drift}
\label{sec:exact}
\subsection{Joint Distribution of Arrival Time and Location}
We first introduce several geometry and drift parameters as $\rho=r_0/r>1$, $\beta=vr/(2D)$, $b=|\beta|$, and $\kappa=v^2/(4D)$, where $\rho$, $\beta$, and $b$ are dimensionless while $\kappa$ is a rate. For $m\geq0$, we set $\nu=m+1/2$ and define the spectral kernel as
\begin{equation}\label{eq:spectral}
k_m(q)=q\frac{J_\nu(\rho q)Y_\nu(q)-J_\nu(q)Y_\nu(\rho q)}
{J_\nu(q)^2+Y_\nu(q)^2},
\end{equation}
where $J_\nu$ and $Y_\nu$ are Bessel functions. Accordingly, the radial time modes are expressed as
\begin{equation}\label{eq:radial}
f_m(t)=-\frac{2D}{\pi r^2\sqrt{\rho}}
\int_0^\infty k_m(q)e^{-Dq^2t/r^2}\dd q,
~~t>0.
\end{equation}
After diffusion and surface-measure scaling, the joint arrival-time and absorption-location law for drift-free diffusion outside a sphere~\cite[Theorem~3.2]{yin2009} leads to the joint time-polar-angle density given by
\begin{equation}\label{eq:g0}
g_0(t,\theta)=\sin\theta\sum_{m=0}^\infty
\left(m+\tfrac12\right)P_m(\cos\theta)f_m(t),
\end{equation}
where $P_m$ is the Legendre polynomial. We clarify that the radial time modes are nonnegative subdensities, generally not normalized to one.

For the finite-time first hit, the known constant-drift change of measure \cite[Theorem~4.2]{yin2009}, also used in \cite{lee2026}, gives
\begin{equation}\label{eq:girsanov}
q_v(t,\boldsymbol{y})=
\exp\!\left(\frac{v\boldsymbol{k}\cdot(\boldsymbol{y}-\boldsymbol{x}_0)}{2D}
-\kappa t\right)q_0(t,\boldsymbol{y}),
\end{equation}
where $\cdot$ is the Euclidean inner product on $\mathbb{R}^3$. We note that evaluating the Brownian exponential likelihood at the stopped path, localized to a finite hitting-time interval, avoids assuming almost-sure absorption. Since $\boldsymbol{k}\cdot\boldsymbol{y}=r\cos\theta$, we obtain
\begin{equation}\label{eq:jointtilt}
g_v(t,\theta)=e^{\beta(\cos\theta-\rho)-\kappa t}
g_0(t,\theta).
\end{equation}
We note that \eqref{eq:jointtilt} specializes the known drift reweighting to this geometry, retaining joint time-location information for cap counting.

\subsection{Cap Impulse and Cumulative Responses}

We now define the drift-weighted angular coefficient and cumulative radial mode by $\mathcal I_m(\beta,\alpha)=\int_{\cos\alpha}^{1}e^{\beta x}P_m(x)\dd x$ and $R_m(t;b)=\int_0^t e^{-\kappa s}f_m(s)\dd s$, respectively. Substituting \eqref{eq:g0} into \eqref{eq:jointtilt} and integrating over the cap, we obtain the exact responses as
\begin{align}\label{eq:maincir}
h_v(t;\alpha)=e^{-\beta\rho-\kappa t}\sum_{m=0}^\infty
\left(m+\tfrac12\right)\mathcal I_m(\beta,\alpha)f_m(t)
\end{align}
and
\begin{align}\label{eq:maincdf}
H_v(t;\alpha)=e^{-\beta\rho}\sum_{m=0}^\infty
\left(m+\tfrac12\right)\mathcal I_m(\beta,\alpha)R_m(t;b).
\end{align}
These exact infinite series require truncation and quadrature for numerical evaluation.

The Laplace transform of each radial time mode \cite[Eq.~(3.9)]{yin2009} solves the exterior modified-Helmholtz problem: its radial factor is $R^{-1/2}K_{m+1/2}(R\sqrt{s/D})$, normalized at $R=r$ and decaying at infinity. At $R=r_0$ and Laplace parameter $s=\kappa$, we obtain the infinite-time cumulative value as
\begin{equation}\label{eq:eventualmodes}
A_m(b)\equiv R_m(\infty;b)=
\begin{cases}
\rho^{-1/2}\dfrac{K_{m+1/2}(\rho b)}{K_{m+1/2}(b)},&b>0,\\[5pt]
\rho^{-m-1},&b=0,
\end{cases}
\end{equation}
where $K_\nu$ is the modified Bessel function of the second kind. Using \eqref{eq:eventualmodes}, we obtain
\begin{equation}\label{eq:eventualcap}
H_v(\infty;\alpha)=e^{-\beta\rho}\sum_{m=0}^\infty
\left(m+\tfrac12\right)\mathcal I_m(\beta,\alpha)A_m(b).
\end{equation}
To justify the limiting operations, let $\mathcal R_s$ be the drift-free radial diffusion and $\tau_r$ its first hitting time of $r$. The radial Feynman-Kac formula \cite{oksendal2003} gives the following representation of $f_m(t)\dd t$ as a nonnegative measure:
\begin{equation}\label{eq:fmt_represent}
f_m(t)\dd t =\mathbb E_{r_0}\left[e^{-Dm(m+1)\int_0^{\tau_r}\mathcal R_s^{-2}\dd s}; \,\tau_r\in\dd t
\right].
\end{equation}
Noting its total mass as $A_m(0)=\rho^{-m-1}$, we find that $0\leq R_m(t;b)\leq A_m(b)\leq\rho^{-m-1}$.
Since $|P_m|\leq1$, $|\mathcal I_m|\leq2e^{|\beta|}$, and $|\partial_\alpha\mathcal I_m|\leq e^{|\beta|}$, the geometric summability for $\rho>1$ justifies time/cap integration and uniform convergence of the cumulative/eventual series and their first aperture derivatives. The same bound applies to $|2R_m-A_m|$, used below.
The CIR representation follows the joint-density expansion for $t>0$. The bounds above control the time-integrated modes $R_m$ and $A_m$; they do not, by themselves, provide a uniform truncation-error bound for the instantaneous CIR.

\subsection{Stable Evaluation of the Cumulative Response}

We note that integrating the spectral CIR from zero to $t$ removes high-frequency exponential damping, leaving a conditionally convergent integral. Instead, we use $R_m(t;b)=A_m(b)-\int_t^\infty e^{-\kappa s}f_m(s)\dd s$ and integrate the damped tail for $t>0$ to obtain
\begin{equation}\label{eq:stablecdf}
R_m(t;b)=A_m(b)+\frac{2}{\pi\sqrt{\rho}}
\int_0^\infty\frac{k_m(q)e^{-(D/r^2)(q^2+b^2)t}}{q^2+b^2}\dd q.
\end{equation}
The integral starts at zero, retains exponential damping at large $q$, and separates eventual-mass accuracy from the finite-time residual. Appendix \ref{app:numerics} details the numerical evaluation of cap responses, including the modal truncation and quadrature settings, and reports convergence checks based on modal and quadrature refinement.

At $\beta=0$, the angular coefficients have the nonsingular limit, given by
\begin{equation}\label{eq:zerodriftangular}
\mathcal I_m(0,\alpha)=
\begin{cases}
1-\cos\alpha,&m=0,\\[2pt]
\dfrac{P_{m-1}(\cos\alpha)-P_{m+1}(\cos\alpha)}{2m+1},&m\geq1.
\end{cases}
\end{equation}
A zero aperture gives zero response. For a full sphere at zero drift, only $m=0$ remains, which helps recover
\begin{equation}\label{eq:classical}
H_0(t;\pi)=\frac{r}{r_0}\erfc\!\left(\frac{r_0-r}{\sqrt{4Dt}}\right)
~~\text{and}~~H_0(\infty;\pi)=\frac{r}{r_0}.
\end{equation}
At nonzero drift, setting $\alpha=\pi$ recovers the aligned full-surface response. The counted probabilities are nondecreasing in $t$ and in $\alpha$, and satisfy $0\leq H_v(t;\alpha)\leq H_v(\infty;\alpha)\leq1$. These properties also check the evaluated finite modal sum.

\begin{figure*}[t]
\centering
\includegraphics[width=.86\textwidth]{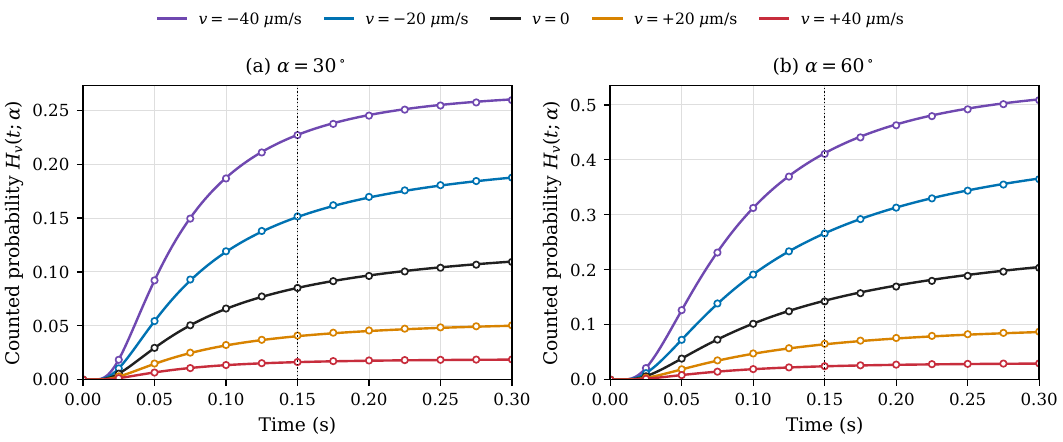}
\caption{Restricted-cap cumulative responses under signed aligned drift. Curves show evaluations of the exact representation, while markers show particle estimates with $250,000$ paths per velocity and $\Delta t=6.25\times10^{-5}$ s. The two counted apertures are $30^\circ$ and $60^\circ$. All trajectories are removed at their first detected global absorption before applying the cap selection.}
\label{fig:validation}
\end{figure*}

\section{SID-Based Aperture Selection}\label{sec:design}

For the symbol duration $T_s$, the counted channel taps are $p_n^{(v)}(\alpha)=H_v(nT_s;\alpha)-H_v((n-1)T_s;\alpha)$.
To distinguish useful molecule capture from late-arriving molecule capture, we define the corresponding counted probabilities as $S_v(\alpha)=H_v(T_s;\alpha)$ and $I_v(\alpha)=H_v(\infty;\alpha)-H_v(T_s;\alpha)$. Then we obtain $S_v=p_1^{(v)}$ and $I_v=\sum_{n\geq2}p_n^{(v)}$. Following the SID criterion of \cite{akdeniz2018}, we further obtain $\SID_v(\alpha)=S_v(\alpha)-I_v(\alpha)=2H_v(T_s;\alpha)-H_v(\infty;\alpha)$.
Finally, we suppress the dependence on $T_s$, unless comparing symbol durations. This is an unnormalized probability difference that balances useful molecule capture and late-arriving molecule capture. It is neither an interference ratio nor a BER, and does not specify a detector or noise model.

For the known drift, we consider unconstrained counting-aperture selection, either offline or by reconfiguring spatially resolved counts. We define the SID-maximizing aperture by $\alpha_v^{\ast}\in\arg\max_{\alpha\in[0,\pi]}\SID_v(\alpha)$. The SID loss incurred by retaining the exact zero-drift optimum for the same symbol duration is
\begin{equation}\label{eq:fixedloss}
L_{\mathrm{fixed}}(v)=\SID_v(\alpha_v^{\ast})-\SID_v(\alpha_0^{\ast}).
\end{equation}
We note from \eqref{eq:fixedloss} that both endpoints are feasible, while $\alpha=0$ records no molecules and gives SID zero, and $\alpha=\pi$ counts absorption events over the entire receiver surface. Specifically, a minimum capture or detector constraint would define a different optimization problem, and no BER-optimality claim is implied. Although \eqref{eq:fixedloss} is nonnegative, the aperture displacement and loss magnitude remain important design quantities.

To examine how drift affects aperture selection, we first introduce the modal coefficient $C_m(b)=2R_m(T_s;b)-A_m(b)$ and define
\begin{equation}\label{eq:stationarymode}
F_b(x)=\sum_{m=0}^\infty\left(m+\tfrac12\right)C_m(b)P_m(x).
\end{equation}
Differentiating the angular integral gives
\begin{equation}\label{eq:sidderivative}
\frac{\partial\SID_v(\alpha)}{\partial\alpha}
=e^{\beta(\cos\alpha-\rho)}\sin\alpha\,F_b(\cos\alpha).
\end{equation}
The derivative balances the useful molecule contribution and late-arriving molecule contribution of an infinitesimal ring. Its positive spatial factor separates the role of drift direction from the timing mechanism below.

To characterize the conditional arrival timing underlying this derivative, we introduce the quantities for an interior angle $\theta\in(0,\pi)$ as
\begin{equation}\label{eq:weightedangularmass}
Z_\kappa(\theta)=\int_0^\infty e^{-\kappa t}g_0(t,\theta)\dd t
\end{equation}
and
\begin{equation}\label{eq:conditionalearly}
\eta_\kappa(\theta;T_s)=\frac{\int_0^{T_s}e^{-\kappa t}g_0(t,\theta)\dd t}
{Z_\kappa(\theta)},
\end{equation}
where $\eta_\kappa$ is the fraction of molecules arriving within the symbol interval, conditioned on eventual arrival at that angle. The signed spatial factor in \eqref{eq:jointtilt} cancels from this conditional ratio.

\begin{proposition}[Conditional timing and counting aperture]\label{prop:timing}
We assume $0<Z_\kappa(\theta)<\infty$ at the interior angles. The fraction $\eta_\kappa(\theta;T_s)$ is nondecreasing in $\kappa\geq0$ for fixed $T_s$, and in $T_s$ for fixed $\kappa$. Opposite drift signs with the same magnitude have identical interior stationary angles and local increasing/decreasing regions of SID. Furthermore, we suppose that for each pair $(\kappa,T_s)$ being compared, there is an $a_\kappa(T_s)\in[0,\pi]$ such that $\eta_\kappa(\theta;T_s)>1/2$ before that angle and $\eta_\kappa(\theta;T_s)<1/2$ after it, within $(0,\pi)$. Then $a_\kappa(T_s)$ is the unique SID-maximizing aperture for either drift sign. Among profiles satisfying this condition, $a_\kappa(T_s)$ is nondecreasing in $\kappa$ at fixed $T_s$, and in $T_s$ at fixed $\kappa$. The proof of this proposition is given in Appendix \ref{app:proof}.
\end{proposition}

Increasing $|v|$ exponentially downweights late times relative to early times \emph{conditional on arrival at an angle}. This does not imply greater absolute capture, which still depends on the drift sign. Consequently, a zero-drift aperture may exclude rings that become useful under either drift direction. The single-crossing condition is stated explicitly: the proposition does not indicate that arbitrary geometries or objectives always possess a unique maximum. Numerically, all stationary roots and both endpoints are compared rather than assuming this condition holds.

\section{Numerical Results}\label{sec:results}

\subsection{Simulation Setup and Response Validation}

We use $D=80\,\um^2$/s, $r=5\,\um$, $r_0=10\,\um$, and initially $T_s=0.15$ s in the simulation. For particle validation, we simulate $N=250{,}000$ trajectories for each velocity $v\in\{-40,-20,0,20,40\}\,\um$/s, up to $0.30$ s. The aperture-design curves are evaluated at $v=-40,-38,\ldots,40\,\um$/s. Settings are fixed before simulation.

Fig. \ref{fig:validation} compares the cap CDFs obtained from the exact representation with particle estimates for both drift signs. The largest discrepancy on the 80-point grid is $1.61\times10^{-3}$. All comparisons meet prespecified tolerances which allow for three sampling standard errors plus a fixed numerical margin. Time-step comparisons, which share Brownian paths, assess detector sensitivity, as described in Appendix \ref{app:numerics}. Global absorption precedes cap filtering, but boundary detection remains approximate. SID always uses the analytical eventual probability \eqref{eq:eventualcap}, but not the simulation horizon.

\subsection{Comparison with Approximate Zero-Drift Model}

At $T_s=0.15$ s, we compare with the normalized approximate zero-drift response from 
\cite{akdeniz2018}. Let $\ell(\theta)=(r_0^2+r^2-2r_0r\cos\theta)^{1/2}$ and $w_t(\theta)=\frac{\sin\theta}{\ell(\theta)^3}\erfc\!\left(\frac{\ell(\theta)}{\sqrt{4Dt}}\right)$.
Then the cumulative cap model is formulated as
\begin{equation}\label{eq:approxbenchmark}
F_A(t;\alpha)=H_0(t;\pi)
\frac{\int_0^\alpha w_t(\theta)\dd\theta}
{\int_0^\pi w_t(\theta)\dd\theta}.
\end{equation}
Optimizing $2F_A(T_s;\alpha)-F_A(\infty;\alpha)$, including the normalized infinite-time limit and endpoints, gives ${27.62}^\circ$, versus the exact ${29.13}^\circ$. Evaluating the approximate-model optimum under the exact response gives an SID of ${0.02529}$, which is $1.64\times10^{-4}$ below the exact maximum of ${0.02545}$. Given such results, our subsequent comparisons use the exact zero-drift optimum to isolate drift mismatch from approximation errors.

\subsection{SID Loss from Retaining the Zero-Drift Aperture}

\begin{figure*}[t]
\centering
\includegraphics[width=.88\textwidth]{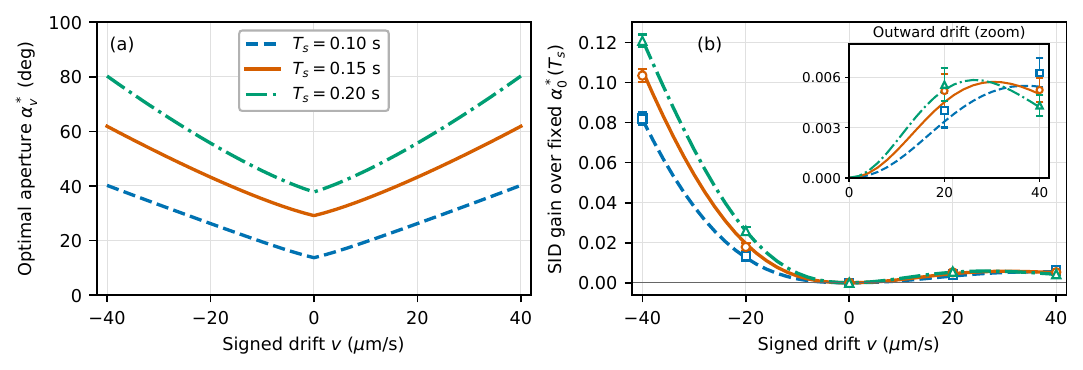}
\caption{Symbol-duration sensitivity of drift-aware counting design at fixed geometry. Left: SID-optimal apertures from the evaluated exact response. Right: absolute SID loss relative to the exact zero-drift optimum recomputed separately for each $T_s$, with an inset for outward drift. Particle markers combine empirical finite-time counts with analytical eventual probabilities [\eqref{eq:hybridgain}, Appendix \ref{app:numerics}]; pointwise 95\% sampling intervals exclude time-step bias. The same trajectories are reused across durations. At $v=0$, each design coincides with its own baseline and thus, the loss is identically zero.}
\label{fig:design}
\end{figure*}

At $T_s=0.15$ s, we quantify the SID loss $L_{\mathrm{fixed}}(v)$ from retaining the exact zero-drift optimal aperture. The SID-maximizing aperture increases from $29.13^\circ$ at zero drift to $43.22^\circ$ and $61.86^\circ$ at $|v|=20$ and $40\,\um$/s, respectively. Opposite drift signs give the same optimal aperture in these cases, but the corresponding losses at $v=-40$ and $+40\,\um$/s are $0.10611$ and $0.00500$, respectively, as shown in Fig. \ref{fig:design} and Table \ref{tab:design}.

We numerically observe a single positive-to-negative crossing of the SID derivative over $0<\alpha<\pi$ for all 41 tested velocities. Conditional timing explains the widening: increasing $|v|$ increases $\kappa$ and exponentially downweights late-arriving molecules relative to early-arriving molecules, conditional on eventual arrival at each interior angle. Hence, $\eta_\kappa(\theta;T_s)$ in \eqref{eq:conditionalearly} is nondecreasing for fixed $T_s$. Under the stated single-crossing condition, \emph{Proposition \ref{prop:timing}} gives a SID-maximizing aperture that is nondecreasing in $|v|$. The aperture sign symmetry follows from \eqref{eq:sidderivative}: the spatial factor is positive over $0<\alpha<\pi$, while $F_b$ depends only on $|v|$. Opposite drift signs therefore have identical interior stationary angles and local increasing/decreasing regions of SID and, under the same condition, the same optimal aperture. Absolute capture depends on the drift sign; thus, SID values and adaptation gains can differ.

\begin{table}[t]
\caption{Aperture adaptation at $T_s=0.15$~s versus its fixed zero-drift design.}
\label{tab:design}
\centering\footnotesize
\setlength{\tabcolsep}{4pt}
\begin{tabular}{@{}rrrrr@{}}
\toprule
$v$ ($\um$/s) & $\alpha_v^*$ (${}^\circ$) & $\SID_v(\alpha_v^*)$ & $\SID_v(\alpha_0^*)$ & $L_{\mathrm{fixed}}$\\
\midrule
$-40$ & 61.86 & 0.28611 & 0.18000 & 0.10611\\
$-20$ & 43.22 & 0.10769 & 0.08818 & 0.01952\\
0 & 29.13 & 0.02545 & 0.02545 & 0\\
$+20$ & 43.22 & 0.02828 & 0.02376 & 0.00452\\
$+40$ & 61.86 & 0.01799 & 0.01299 & 0.00500\\
\bottomrule
\end{tabular}
\end{table}

For outward drift at $v=+40\,\um$/s, widening the cap from $29.13^\circ$ to $61.86^\circ$ increases the useful molecule probability $S_v$ from $0.01573$ to $0.02436$ and the late-arriving molecule probability $I_v$ from $0.00274$ to $0.00637$. The added useful molecule probability, $0.00863$, exceeds the added late-arriving molecule probability, $0.00363$, giving an SID gain of $0.00500$.

We assess the finite-time component of the predicted gains using paired particle counts combined with analytical eventual probabilities, as in \eqref{eq:hybridgain} in Appendix \ref{app:numerics}. All four nonzero-drift gain estimates have positive pointwise 95\% sampling intervals. For example, at $v=+40\,\um$/s, the estimated gain is $0.00525$ with interval $[0.00452,0.00599]$. These intervals quantify sampling variation under the discretized detector, excluding time-step bias, and do not independently validate infinite-horizon SID.

\subsection{Sensitivity to Symbol Duration}

For $T_s=0.10$, $0.15$, and $0.20$~s at the same geometry and 41 signed velocities, each duration uses its own exact zero-drift optimum $\alpha_0^*(T_s)$. Thus, $L_{\mathrm{fixed}}(v;T_s)$ isolates drift mismatch. Reused first-hit records give correlated comparisons.

The optimal apertures for $T_s=0.10$, $0.15$, and $0.20$~s are $13.72^\circ$, $29.13^\circ$, and $37.79^\circ$ at zero drift, respectively, but become $40.21^\circ$, $61.86^\circ$, and $80.26^\circ$ at $|v|=40\,\um$/s, as shown in Fig. \ref{fig:design}. Endpoint and higher-order search confirm the single positive-to-negative crossing of \emph{Proposition \ref{prop:timing}} for all 123 evaluated pairs. Longer intervals classify more arriving molecules as useful, allowing additional rings to pass the threshold. This crossing evidence is numerical and grid-specific. Moreover, for $T_s=0.10$, $0.15$, and $0.20$ s, the losses from retaining each duration's zero-drift optimum are $0.08173$, $0.10611$, and $0.12143$ at $v=-40\,\um$/s, respectively, while the corresponding losses at $v=+40\,\um$/s are $0.00545$, $0.00500$, and $0.00422$, respectively. The optimal aperture is determined by the marginal balance between useful molecules and late-arriving molecules. The resulting SID gain also depends on the capture probabilities over the added angular region and the zero-drift reference aperture for each duration.

All 12 nonzero-drift gain estimates relative to each duration's zero-drift optimum have positive pointwise sampling intervals. Appendix \ref{app:numerics} reports CDF checks at the design times and apertures and gain comparisons between the two smallest steps, which assess sensitivity without bounding residual detector bias. In a secondary comparison with full-sphere counting, the smaller SID gains at $v=+40\,\um$/s remain unresolved for $T_s=0.15$ and $0.20$ s. These results concern the tested geometry and the SID objective; they do not establish BER or capacity improvements.

Overall, the tested cases show that the SID-maximizing aperture widens with drift magnitude and symbol duration, while the loss from retaining the zero-drift aperture differs markedly between drift directions. These results support the importance of selecting the counting aperture according to the operating drift and symbol duration.

\section{Conclusion}

In this work, we developed exact cap-response representations that connect the known joint hitting law to aperture selection under aligned drift. Our analysis distinguishes conditional arrival timing from absolute capture, revealing that under the stated single-crossing condition, opposite drifts of equal magnitude can favor the same aperture while yielding different SID payoffs. Numerical results further showed that retaining each duration's zero-drift aperture incurs SID loss in the tested nonzero-drift cases. Within the fully absorbing counting model and the SID objective, these findings clarified the tradeoff between useful molecule capture and ISI, providing a basis for aperture selection as per the operating drift and symbol duration. 
\appendices
\section{Numerical Implementation and Validation}
\label{app:numerics}

\subsection{Response Evaluation and Aperture Optimization}

For response evaluation, we truncate the modal series at $m=40$ and use 20-point Gauss-Legendre quadrature on each spectral $q$-panel. The panel endpoints are $0,0.0025,0.005,\ldots,0.05$, $0.10,0.15,\ldots,1$, and $2,\ldots,60$. Angular integration uses adaptive quadrature for responses and 64-point Gauss-Legendre quadrature for design. Modal/quadrature refinement, probability bounds, cap nesting, and time monotonicity are checked. Independent adaptive quadrature through mode $m=60$ agrees on the 80-point grid below within $1.4\times10^{-13}$; four inverse-Laplace checks agree within $7.4\times10^{-14}$. The tested dimensionless region is $\rho=2$, $|\beta|\leq1.25$, and $DT_s/r^2\in\{0.32,0.48,0.64\}$; these checks are not uniform error guarantees.

To identify the SID-maximizing aperture, we compare the SID values at all admissible stationary candidates obtained from the truncated polynomial in~\eqref{eq:stationarymode} and at both endpoints, $\alpha=0$ and $\alpha=\pi$. Higher-order independent optimization and a $0.05^\circ$ grid confirm the maxima. SID uses~\eqref{eq:eventualcap} for eventual mass. At very small times, zero is returned only if a tangent-plane first-passage bound is below $10^{-14}$, while negative values are not clipped. Direct angular quadrature avoids forward-recurrence cancellation near zero drift.

\subsection{Particle Simulation and Absorption Detection}

Free propagation uses exact Gaussian drift-diffusion increments at three nested steps, i.e., $\Delta t=2.5\times10^{-4}$, $1.25\times10^{-4}$, and $6.25\times10^{-5}$ s. For each velocity, simulations at the three time-step sizes share the same underlying Brownian paths, enabling paired comparisons of the absorption estimates. PCG64 seeds 2026092601 through 2026092605 correspond to ascending velocity order.

The detector shifts the radius to $r+0.582597\sqrt{2D\Delta t}$ for reducing missed crossings~\cite{gobet2010}. At the first detected absorption, the endpoint direction determines whether the hit is counted, while the trajectory is always removed. This approximates continuous first contact. Its remaining cap-indicator error is assessed by nested-step comparisons, and a smooth-functional boundary-correction result is not assumed to apply unchanged.


\subsection{Validation Criteria and Sampling Uncertainty}

We assess particle estimates of the CDFs and SID gains through comparisons with analytical predictions and paired comparisons between time steps. For these comparisons, we denote the marginal binomial standard error by $\mathrm{SE}$ and the standard error of the corresponding paired estimator computed from shared trajectories by $\mathrm{SE}_{\mathrm{paired}}$.

The 80-point validation grid uses times $0.05$, $0.10$, $0.15$, and $0.30$ s and apertures $30^\circ$, $60^\circ$, $180^\circ$, and $\alpha_v^{\ast}$ for each velocity, where $\alpha_v^{\ast}$ is optimized at the baseline $T_s=0.15$ s. All CDF errors, where the maximum is $1.61\times10^{-3}$, satisfy the prespecified tolerance $3\mathrm{SE}+5\times10^{-4}$. All 80 paired differences between the two smallest time steps satisfy $3\mathrm{SE}_{\mathrm{paired}}+2.5\times10^{-4}$, and the largest paired difference between these two steps is $2.92\times10^{-4}$. Each tolerance allows for three sampling standard errors plus a fixed numerical margin. They serve as numerical validation criteria and do not guarantee simultaneous confidence coverage or vanishing discretization bias.

For aperture comparisons based on shared trajectories, we denote $\widehat{\tau}_i$ and $\widehat{\Theta}_i$ as the detected first-hit time and polar angle of trajectory $i$, respectively, and define $Z_i(\alpha)=\ind\{\widehat{\tau}_i\leq T_s,\widehat{\Theta}_i\leq\alpha\}$. Combining empirical finite-time counts with analytical eventual probabilities, we express the gain estimator as
\begin{align}\label{eq:hybridgain}
\widehat L_{\mathrm{hyb}}(v)
=&\frac{2}{N}\sum_{i=1}^{N}\bigl[Z_i(\alpha_v^{\ast})-Z_i(\alpha_0^{\ast})\bigr]\notag\\
&-\bigl[H_v(\infty;\alpha_v^{\ast})-H_v(\infty;\alpha_0^{\ast})\bigr].
\end{align}
For comparisons with full-sphere counting, we replace $\alpha_0^{\ast}$ by $\pi$ in~\eqref{eq:hybridgain}. For nested caps, the sum counts hits in their intervening spherical surface band, signed by subtraction order. The unsigned count is binomial. Its Clopper-Pearson probability interval transforms affinely into the gain interval, conditional on the analytical eventual difference. The interval covers sampling variation under the discretized detector, excluding time-step bias.

For three durations, 45 CDF cases compare drift-optimal apertures, each duration's zero-drift optimum, and full-sphere counting. All errors at the smallest step satisfy $3\mathrm{SE}+0.0005$ and all 45 paired changes between the two smallest steps satisfy $3\mathrm{SE}_{\mathrm{paired}}+0.00025$, while their maxima are $1.69\times10^{-3}$ and $3.60\times10^{-4}$, respectively. We evaluate 30 SID gain estimates across three durations and five velocities, using each duration's exact zero-drift optimum and full-sphere counting as two baselines. All theory comparisons satisfy $3\mathrm{SE}_{\mathrm{paired}}+0.001$ and all 30 changes between these steps satisfy $3\mathrm{SE}_{\mathrm{paired}}+0.0005$, while the largest paired gain change is $8.32\times10^{-4}$. These correlated comparisons include structural zeros, such that they assess sensitivity at the design times and apertures, without bounding residual detector bias. 
\section{Proof of Proposition \ref{prop:timing}}\label{app:proof}

For a fixed interior angle $\theta$, we denote the integrals defining $Z_\kappa(\theta)$ in~\eqref{eq:weightedangularmass} over $[0,T_s]$ and $(T_s,\infty)$ by $A_\kappa$ and $B_\kappa$, respectively. Thus, $Z_\kappa(\theta)=A_\kappa+B_\kappa$ and $\eta_\kappa(\theta;T_s)=A_\kappa/Z_\kappa(\theta)$.

To establish monotonicity in $\kappa$, let $\kappa_2=\kappa_1+\delta$ with $\delta\geq0$. We then obtain
\begin{align}\label{eq:tiltproof}
A_{\kappa_2}B_{\kappa_1}-A_{\kappa_1}B_{\kappa_2}
=&\int_0^{T_s}\!\int_{T_s}^{\infty}e^{-\kappa_1(s+t)}(e^{-\delta s}-e^{-\delta t})\nonumber\\
&\hspace{0em}\times g_0(s,\theta)g_0(t,\theta)\dd t\dd s\ \geq0,
\end{align}
where the inequality is due to $s\leq T_s<t$ and the nonnegativity of $g_0$. Dividing \eqref{eq:tiltproof} by $Z_{\kappa_1}(\theta)Z_{\kappa_2}(\theta)>0$ gives $\eta_{\kappa_2}(\theta;T_s)\geq\eta_{\kappa_1}(\theta;T_s)$, without requiring any hitting-time moment assumption. For fixed $\kappa$, monotonicity in $T_s$ follows because enlarging the integration interval cannot decrease the numerator, while $Z_\kappa(\theta)$ remains unchanged.

To connect these timing properties to aperture selection, we use \eqref{eq:jointtilt} to express the SID derivative as
\begin{equation}\label{eq:ringcriterion}
\partial_\alpha\SID_v(\alpha)
=e^{\beta(\cos\alpha-\rho)}Z_\kappa(\alpha)
\left[2\eta_\kappa(\alpha;T_s)-1\right].
\end{equation}
The prefactor is positive at the interior angles under consideration, and $\eta_\kappa$ depends on the drift only through $\kappa=v^2/(4D)$. Hence, opposite drifts of equal magnitude have identical interior stationary angles and identical regions in which SID increases or decreases.

Under the stated single-crossing condition, we see that SID strictly increases before $a_\kappa(T_s)$ and strictly decreases after it. Therefore, $a_\kappa(T_s)$ is the unique global SID-maximizing aperture for either drift sign.

Finally, the pointwise monotonicity of $\eta_\kappa$ implies the angular regions where $\eta_\kappa(\theta;T_s)>1/2$ can only expand as either $\kappa$ or $T_s$ increases with the other held fixed. Under the single-crossing condition, these regions determine $a_\kappa(T_s)$. Thus, the maximizing aperture is nondecreasing in each parameter when the other is fixed.

\section*{Acknowledgment}
The authors used ChatGPT (OpenAI; GPT-6 Astra) as an assistive tool for language refinement, manuscript organization, and routine debugging of selected computational scripts. All mathematical derivations, numerical methodologies, reported results, scientific interpretations, and conclusions were independently verified and approved by the authors.


\begin{thebibliography}{00}
\bibitem{akyildiz2008}
I.~F. Akyildiz, F.~Brunetti, and C.~Bl\'{a}zquez, ``Nanonetworks: A new communication paradigm,'' \emph{Computer Networks}, vol.~52, no.~12, pp.~2260--2279, 2008.

\bibitem{farsad2016}
N.~Farsad, H.~B. Yilmaz, A.~Eckford, C.-B. Chae, and W.~Guo, ``A comprehensive survey of recent advancements in molecular communication,'' \emph{IEEE Commun. Surveys Tuts.}, vol.~18, no.~3, pp.~1887--1919, 2016.

\bibitem{jamali2019}
V.~Jamali, A.~Ahmadzadeh, W.~Wicke, A.~Noel, and R.~Schober, ``Channel modeling for diffusive molecular communication---a tutorial review,'' \emph{Proc. IEEE}, vol.~107, no.~7, pp.~1256--1301, 2019.

\bibitem{yeh2012new}
P.-C. Yeh, K.-C. Chen, Y.-C. Lee, L.-S. Meng, P.-J. Shih, P.-Y. Ko, W.-A. Lin, and C.-H. Lee, ``A new frontier of wireless communication theory: Diffusion-based molecular communications,'' \emph{IEEE Wireless Commun.}, vol.~19, no.~5, pp.~28--35, 2012.

\bibitem{noel2014flows}
A.~Noel, K.~C. Cheung, and R.~Schober, ``Diffusive molecular communication with disruptive flows,'' in \emph{Proc. IEEE Int. Conf. Commun. (ICC)}, Sydney, Australia, Jun. 2014, pp.~3600--3606.

\bibitem{akdeniz2018}
B.~C. Akdeniz, N.~A. Turgut, H.~B. Yilmaz, C.-B. Chae, T.~Tugcu, and A.~E. Pusane, ``Molecular signal modeling of a partially counting absorbing spherical receiver,'' \emph{IEEE Trans. Commun.}, vol.~66, no.~12, pp.~6237--6246, 2018.

\bibitem{yilmaz2014}
H.~B. Yilmaz, A.~C. Heren, T.~Tugcu, and C.-B. Chae, ``Three-dimensional channel characteristics for molecular communications with an absorbing receiver,'' \emph{IEEE Commun. Lett.}, vol.~18, no.~6, pp.~929--932, 2014.

\bibitem{lee2026}
Y.-C. Lee, P.-C. Yeh, and C.-H. Lee, ``Exact 3-D channel impulse response under uniform drift for absorbing spherical receivers,'' \emph{IEEE Commun. Lett.}, vol.~30, pp.~1290--1294, 2026.

\bibitem{yin2009}
C.~Yin and C.~Wang, ``Hitting time and place of Brownian motion with drift,'' \emph{Open Stat. Probab. J.}, vol.~1, pp.~38--42, 2009.

\bibitem{huang2023}
X.~Huang, Y.~Fang, S.~T. Johnston, M.~Faria, N.~Yang, and R.~Schober, ``Analysis of MC systems employing receivers covered by heterogeneous receptors,'' \emph{IEEE Trans. Mol. Biol. Multi-Scale Commun.}, vol.~9, no.~1, pp.~63--78, Mar. 2023.

\bibitem{huang2025}
X.~Huang, Y.~Huang, M.~Wen, N.~Yang, and R.~Schober, ``Heterogeneous receptors-based molecule harvesting in MC: Analysis for ISI mitigation and energy efficiency,'' \emph{IEEE Trans. Commun.}, vol.~73, no.~3, pp.~1621--1637, Mar. 2025.

\bibitem{oksendal2003}
B.~\O ksendal, \emph{Stochastic Differential Equations: An Introduction with Applications}, 6th~ed. Springer, 2003.

\bibitem{gobet2010}
E.~Gobet and S.~Menozzi, ``Stopped diffusion processes: Boundary corrections and overshoot,'' \emph{Stochastic Processes and their Applications}, vol.~120, no.~2, pp.~130--162, 2010.
\end{thebibliography}
\end{document}